\documentclass[11pt]{article}
\usepackage{cospar}
\usepackage{url}

\usepackage{graphicx}
\usepackage[figuresright]{rotating}

\newcommand{\EQ}{\begin{equation}}
\newcommand{\EN}{\end{equation}}
\newcommand{\EQA}{\begin{eqnarray}}
\newcommand{\ENA}{\end{eqnarray}}

\newcommand{\Eq}[1]{equation~(\ref{#1})}
\newcommand{\Eqs}[2]{equations~(\ref{#1}) and~(\ref{#2})}

\newcommand{\Fig}[1]{Figure~\ref{#1}}

\newcommand{\Tab}[1]{Table~\ref{#1}}

\newcommand{\bra}[1]{\langle #1\rangle}

\newcommand{\meanS}{\overline{S}}
\newcommand{\meanA}{\overline{A}}
\newcommand{\meanB}{\overline{B}}

\newcommand{\meanAA}{\overline{\mbox{\boldmath $A$}}}
\newcommand{\meanBB}{\overline{\mbox{\boldmath $B$}}}
\newcommand{\meanJJ}{\overline{\mbox{\boldmath $J$}}}

\newcommand{\pp}{\hat{\mbox{\boldmath $\phi$}} {}}

\newcommand{\rrr}{\mbox{\boldmath $r$} {}}

\newcommand{\uu}{\mbox{\boldmath $u$} {}}
\newcommand{\vv}{\mbox{\boldmath $v$} {}}

\newcommand{\bb}{\mbox{\boldmath $b$} {}}

\newcommand{\BB}{\mbox{\boldmath $B$} {}}

\newcommand{\AAA}{\mbox{\boldmath $A$} {}}
\newcommand{\aaa}{\mbox{\boldmath $a$} {}}
\newcommand{\jj}{\mbox{\boldmath $j$} {}}

\newcommand{\SSS}{\mbox{\boldmath $S$} {}}

\newcommand{\nab}{\mbox{\boldmath $\nabla$} {}}

\newcommand{\dd}{{\rm d} {}}

\newcommand{\km}{\,{\rm km}}

\newcommand{\yr}{\,{\rm yr}}

\newcommand{\yan}[5]{, ``#5,'' {\em Astron.\ Nachr.\ }{\bf #2}, #3-#4, #1.}

\newcommand{\yana}[5]{, ``#5,'' {\em Astron.\ Astrophys.\ }{\bf #2}, #3-#4, #1.}
\newcommand{\yanas}[5]{, ``#5,'' {\em Astron.\ Astrophys.\ Suppl.\ }{\bf #2}, #3-#4, #1.}

\newcommand{\yass}[5]{, ``#5,'' {\em Astrophys.\ Spa.\ Sci.\ }{\bf #2}, #3-#4, #1.}

\newcommand{\ysph}[5]{, ``#5,'' {\em Solar Phys.\ }{\bf #2}, #3-#4, #1.}

\newcommand{\ymn}[5]{, ``#5,'' {\em Monthly Notices Roy.\ Astron.\ Soc.\ }{\bf #2}, #3-#4, #1.}

\newcommand{\ynat}[5]{, ``#5,'' {\em Nature }{\bf #2}, #3-#4, #1.}

\newcommand{\yjfm}[5]{, ``#5,'' {\em J.\ Fluid Mech.\ }{\bf #2}, #3-#4, #1.}

\newcommand{\yprl}[5]{, ``#5,'' {\em Phys.\ Rev.\ Letters }{\bf #2}, #3-#4, #1.}

\newcommand{\yapj}[5]{, ``#5,'' {\em Astrophys.\ J.\ }{\bf #2}, #3-#4, #1.}

\newcommand{\ygafd}[5]{, ``#5,'' {\em Geophys.\ Astrophys.\ Fluid Dynam.}{\bf #2}, #3-#4, #1.}

\newcommand{\yjour}[6]{, ``#6,'' {\em #2} {\bf #3}, #4-#5, #1.}

\newcommand{\pproc}[5]{, ``#2,'' In {\em #3} (ed.\ #4), #5, #1 (to appear).}
\newcommand{\yproc}[7]{, ``#4,'' In {\em #5} (ed.\ #6), pp.\ #2-#3.\ #7, #1.}

\newcommand{\ppapjl}[2]{ ``#2,'' {\em Astrophys.\ J.\ Lett.\ } (in press) ~#1.}

\title{How magnetic helicity ejection helps  large scale dynamos}

\author{
A. Brandenburg\address{NORDITA, Blegdamsvej 17, DK-2100 Copenhagen \O, Denmark},
E.\ G. Blackman\address{Department of Physics \& Astronomy,
University of Rochester, Rochester NY 14627, USA},
and
G.\ R.\ Sarson\address{School of Mathematics \& Statistics, University of
Newcastle, Newcastle NE1 7RU, U.K.}
}
\begin{document}

% typeset front matter
\maketitle

\begin{abstract}
There is mounting evidence that the ejection of magnetic helicity from
the solar surface is important for the solar dynamo.
Observations suggest that in the northern hemisphere the magnetic
helicity flux is negative. We propose that this magnetic helicity flux
is mostly due to small scale magnetic fields; in contrast to the 
more systematic large scale field of the 11 year cycle, whose
helicity flux may be of opposite sign, and may be excluded from the
observational interpretation. Using
idealized simulations of MHD turbulence as well as a simple two-scale
model, we show that shedding small scale (helical) field has two important effects.
(i) The strength of the large scale field reaches the observed levels.
(ii) The evolution of the large scale field proceeds on time scales
shorter than the resistive time scale, as would otherwise be enforced
by magnetic helicity conservation.
In other words, the losses ensure that the solar dynamo is always in
the near-kinematic regime.
This requires, however, that the ratio of small scale to large scale
losses cannot be too small, for otherwise the large scale field in the
near-kinematic regime will not reach the observed values.
\end{abstract}

\section*{INTRODUCTION}

In recent years a new term has entered the solar physics vocabulary:
magnetic helicity.
This topic has received significant attention on two quite
different fronts that now seem to converge in their mutual importance.
The detection of helical features has received interest firstly as a purely
diagnostic tool to characterize topological complexity.
As magnetic helicity is a conserved quantity (neglecting boundary losses and 
resistivity), this has secondly been seen
to impose constraints on mean-field dynamo theory;
with the advent of powerful massively parallel computers,
these constraints are now also being confirmed numerically as
progressively larger magnetic Reynolds numbers are becoming possible.

In the following we discuss briefly the observational aspects relevant
to the dynamo problem and turn then to the connection with  dynamo theory.
We then present arguments linking the loss of small scale (SS) field
to an enhancement of the large scale (LS) dynamo.

\section*{THE OBSERVED LARGE SCALE FIELD}

The question of scales is very important: what is large scale to an
observer could be small scale to a dynamo theorist, for example.
Looking at the magnetic field in helmet streamers reveals structures
comparable to the solar radius, 
which might be taken to 
belong to the large scale field.
This may be misleading, however.
For a dynamo theorist who wants to explain the 11 year solar cycle it
matters that there are bipolar regions with systematic orientation and
tilt, but longitudinal departures from the mean axisymmetric form 
are secondary to the underlying effect that sustains the field.
Longitudinal averages are therefore a sensible tool to extract what
a dynamo theorist might want to call large scale field.

That such a definition makes some sense can be seen by looking at
longitudinally averaged magnetograms of the solar surface 
as a function of latitude and time.
We reconstruct such a space-time diagram from the amplitude and phase
coefficients, $\hat{B}_\ell$ and $\phi_\ell$ respectively, given by
Stenflo (1988).
Following earlier work (Stenflo \& Vogel 1986, Stenflo \& G\"udel 1988)
he described the radial component of the longitudinally averaged field
of dipole symmetry in the form
\EQ
\meanB_r(\theta,t)=\sum_{\ell=1,3,...}\hat{B}_\ell
P_\ell(\cos\theta)\,\cos[\omega(t-t_0)+\phi_\ell],
\label{Br}
\EN
%GRS: the following line perhaps unnecessary.
%GRS: NB: we never mention \omega -- is it just fixed at 2 pi/ 11 yrs,
% or was it part of Stenflo's fit to the observations?
%AB: yes, omega is actually also determined, but it comes out close
%AB: to 22 years.
where $P_{\ell}$ is the Legendre polynomial
and $\omega=2\pi/22{\rm years}$ is the cycle frequency;
the values of $\hat{B}_\ell$ and $\phi_\ell$ are reproduced in \Tab{Tstenflo}.

\begin{table}[tb]\caption{
Phase and amplitude coefficients describing the mean dipole symmetry 
radial field at the solar surface, from Stenflo (1988).
The last row gives the contributions to the root-mean-square value
of the field, $\hat{B}_\ell/\sqrt{2(2\ell+1)}$.
The phase is defined relative to the epoch $t_0=1960\yr$.
}\vspace{12pt}\centerline{\begin{tabular}{l|ccccccc}
$\ell$  &   1   &  3   &  5   &  7   &  9   & 11   & 13   \\
\hline
$\phi_\ell/2\pi$
        &  0.24 & 0.29 & 1.07 & 1.59 & 1.95 & 2.45 & 2.85 \\
$\hat{B}_\ell\,\mbox{[gauss]}$
        &  1.49 & 1.66 & 3.15 & 2.28 & 2.64 & 1.56 & 0.92 \\
$\hat{B}_\ell/\sqrt{2(2\ell+1)}\,\mbox{[gauss]}$
        &  0.61 & 0.44 & 0.67 & 0.42 & 0.43 & 0.23 &~0.12
\label{Tstenflo}\end{tabular}}\end{table}

\begin{figure}[t!]\begin{center}
\includegraphics[width=.9\textwidth]{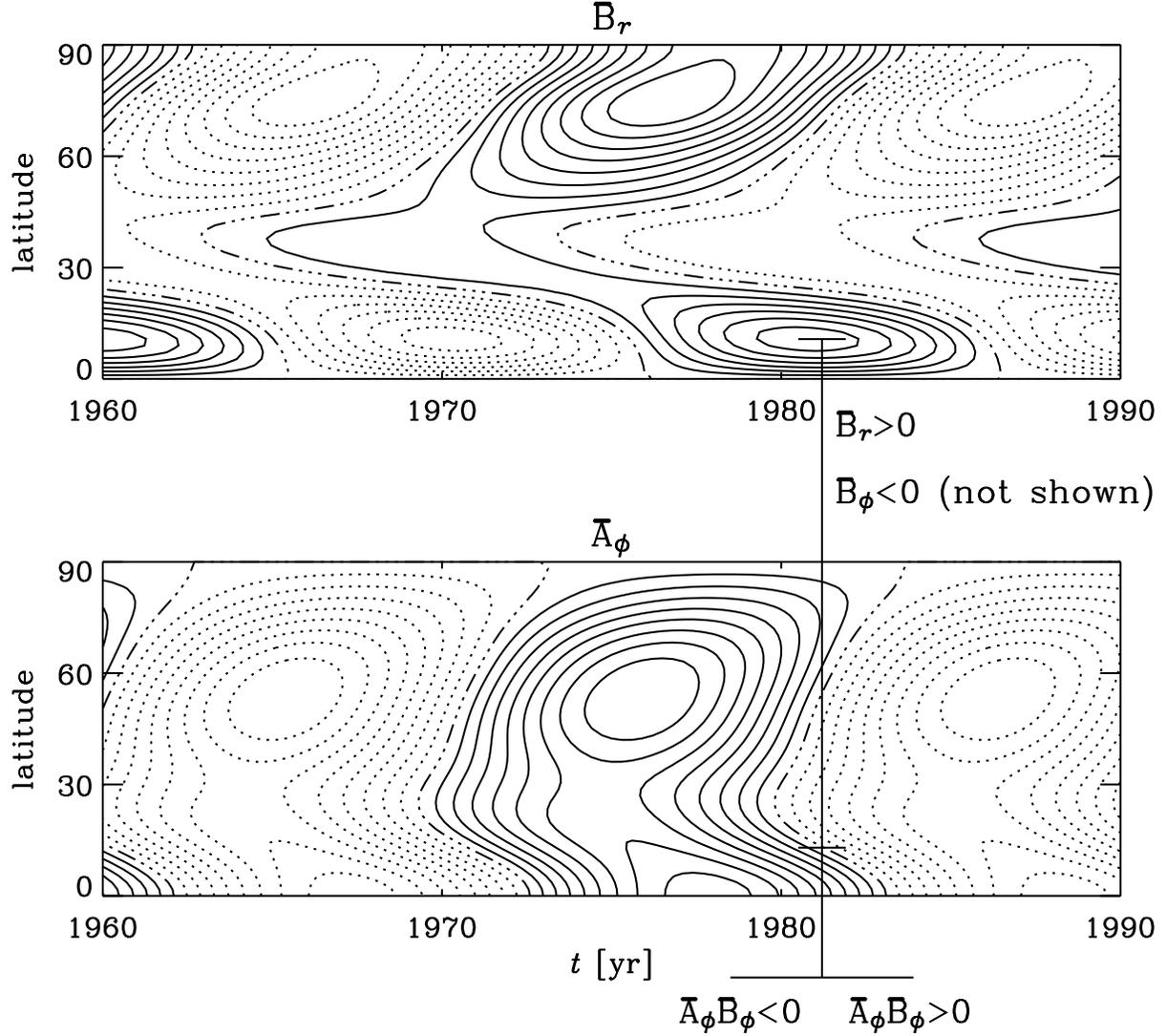}
\end{center}\caption[]{Mean dipole symmetry radial field, $\meanB_r$, 
reconstructed from the coefficients of Stenflo, 1988 (upper panel).
The corresponding toroidal component of the mean vector potential, 
$\meanA_\phi$, derived from $\meanB_r$ (lower panel).
Solid contours denote positive values, dotted contours negative values.
The solar cycle maximum of 1981 is highlighted, 
as is the latitude of 10$^\circ$ where $\meanB_r$ was then strongest.
The signs of various quantities at or around this epoch are
also shown (see text for more details).
}\label{stenflo}\end{figure}

Since the mean field is (by definition) axisymmetric, its poloidal part
can be written in the form $\meanBB_{\rm pol}=\nab\times(\meanA_\phi\pp)$.
This allows us to relate the surface values of $\meanB_r$ to $\meanA_\phi$,
\EQ
\meanB_r={1\over r\sin\theta}\,{\partial\over\partial\theta}\,
\left(\sin\theta\meanA_\phi\right),\quad r=R_\odot.
\EN
Note that no radial derivatives enter, 
so that we can obtain $\meanA_\phi$ by surface integration.
To express $\meanA_\phi$ more simply, 
note that we can also write $\meanBB_{\rm pol}$
in terms of the poloidal potential $\meanS$, i.e.\
\EQ
r\meanB_r=\rrr\cdot\nab\times\nab\times(\meanS\rrr)\equiv-L^2\meanS,
\EN
where $L^2$ is the angular part of the Laplacian,
in spectral space satisfying $-L^2=\ell(\ell+1)$.
Since $\meanA_\phi=-\partial \meanS/\partial\theta$
and $\partial P_\ell/\partial\theta=P_\ell^1$, we have
\EQ
\meanA_\phi(\theta,t)=-\sum_{\ell=1,3,...}{r\over\ell(\ell+1)}\hat{B}_\ell
P_\ell^1(\cos\theta)\,\cos[\omega(t-t_0)+\phi_\ell].
\label{Ap}
\EN
In \Fig{stenflo} we show both $\meanB_r$ and $\meanA_\phi$, as reconstructed
using \Eqs{Br}{Ap}.

Stenflo's coefficients suggest that the sun's large scale magnetic field
is dominated by the higher harmonics around $\ell=5$.
The coefficient of the basic dipole mode, $\hat{B}_1$, is only
the sixth-largest among all coefficients.
This misrepresents the energy present in the various modes, however.
The integral of $\meanB_r^2$ over the full surface and over one cycle
divided by the surface area and the length of the cycle is equal to the
sum of $\hat{B}_\ell^2/[2(2\ell+1)]$.
In \Tab{Tstenflo} its square root, i.e.\ the contribution to the
rms field strength, is given.
Although the contribution from $\ell=5$ is still the largest,
the basic dipole contribution is now the second largest.
In that sense one is justified in talking of the Sun's field as dipolar.

The magnetic helicity of the large scale field, 
as defined above, is an important quantity.
Magnetic helicity, $\int\meanAA\cdot\meanBB\,\dd V$, can be tricky
to work with, since it involves the magnetic vector potential, $\meanAA$
(with $\meanBB=\nab\times\meanAA$), which is not directly observable.
Attempts to reconstruct $\meanAA$ from $\meanBB$ raise the question of
the choice of a gauge;  and the magnetic helicity is not in general 
gauge invariant.
This problem can be avoided by considering instead the gauge invariant magnetic
helicity of Berger \& Field (1984),
\EQ
H=\int_V(\meanAA+\meanAA_{\rm P})
\cdot(\meanBB-\meanBB_{\rm P})\;\dd V,
\label{BF84integral}
\EN
where $\meanBB_{\rm P}=\nab\times\meanAA_{\rm P}$ is a potential reference field
inside $V$ with the boundary condition
$\rrr\cdot\meanBB_{\rm P}=\rrr\cdot\meanBB$.
As shown in Brandenburg, Dobler \& Subramanian (2002), for an
axisymmetric mean field expressed as
$\meanBB=\meanB_\phi\pp+\nab\times(\meanA_\phi\pp)$, with
$\meanA_\phi=\meanB_\phi=0$ on the axis, this yields simply\footnote{For
comparison, we note that in the Coulomb gauge,
$\int\meanAA\cdot\meanBB\,\dd V=2\int\meanA_\phi\meanB_\phi\;\dd V
+\oint(\meanA_\phi\pp\times\meanAA)\cdot\dd\SSS$, which is different from
the gauge invariant magnetic helicity, $H$, of Berger \& Field (1984).}
\EQ
H=2\int_V\meanA_\phi\meanB_\phi\;\dd V.
\label{Haxi}
\EN
There is no way to determine this quantity for the sun without
knowing both $\meanA_\phi$ and $\meanB_\phi$ throughout the entire sphere.
Experience with axisymmetric mean field dynamos shows, however, 
that the product $\meanA_\phi\meanB_\phi$ does not vary strongly in radius,
so considering this quantity at the surface may still be useful.
%GRS: is there a problem here?  is B_phi not almost zero at the surface,
% where the conductivity becomes very low?  it might be better to rephrase
% the above
%AB: but sufficiently below the surface B_phi is probably ok.

The question of the sign of $\meanA_\phi\meanB_\phi$ is related to the
question of the phase relation between toroidal and poloidal field,
which was investigated in the mid seventies (Stix 1976a,b, Yoshimura 1976).
By estimating the phase shift between $\meanB_r$ and $\meanB_\phi$,
it was argued that the radial angular velocity gradient,
$\partial\Omega/\partial r$, should be negative.
This, together with the fact that the solar dynamo wave propagates
equatorward, suggested that $\alpha$ is positive (negative) in the
northern (southern) hemisphere.
Our knowledge has changed since then: 
(i) we now know from helioseismology that
$\partial\Omega/\partial r$ is positive (suggesting that something
%GRS: i changed the next word from *is* to *may be*,  OK?
% (or is this not free from additional complications.)
%AB: yes, right.
may be wrong with the observed phase relation); 
(ii) the direction of
propagation of the dynamo wave can be reversed by meridional circulation
(Durney 1995, Choudhuri, Sch\"ussler \& Dikpati 1995).
The question of the sign of $\meanA_\phi\meanB_\phi$ differs somewhat 
from that of the phase relation;
the former should depend on the sign of $\alpha$, not on the sign of
$\partial\Omega/\partial r$.
Let us ignore for a moment the problems with the
traditional phase relation then, and discuss what can be said about the
sign of $\meanA_\phi\meanB_\phi$ at the surface.

Yoshimura (1976) estimated the magnitude of the toroidal field by
averaging the unsigned surface flux in an appropriate fashion.
The idea is that the toroidal field emerges as $\Omega$-loops at
the surface, and that a stronger toroidal field shows up as an increased
level of unsigned flux, i.e.\ $|\meanB_\phi|\propto\overline{|B_r|}$
(note that the modulus is underneath the average of $B_r$).
He then determined the sign of $\meanB_\phi$ simply by looking at the
orientation of the bipolar regions on solar magnetograms.
This tells us that during the maximum of Cycle 21
in 1981 or 1982 the toroidal field was negative (positive)
in the northern (southern) hemisphere.\footnote{See
\url{http://www.hao.ucar.edu/public/education/slides/slide19.html}}
Comparing with $\meanB_r$ of \Fig{stenflo} (upper panel) we see that
$\meanB_r$ was positive (solid contours at 10$^\circ$;
here we focus on latitudes near the equator where $\meanB_r$ is strongest);
hence $\meanB_r\meanB_\phi$ was negative.
This verifies the original finding of Stix (1976a) and Yoshimura (1976).
%GRS: What epoch did Stix and Yoshimura study?  Or was this quantity of
% fixed sign throughout the cycle.
%AB: some 15 years maybe or less. This statement is only restricted
%AB: to lower latitudes where the toroidal field is strong.
Comparing with $\meanA_\phi$ of \Fig{stenflo} (lower panel),
the situation is less clear;
$\meanA_\phi\meanB_\phi$ may be negative
just before solar maximum and positive just after solar maximum.
This suggests that the gauge invariant magnetic helicity $H$
(given by \Eq{Haxi}) might be close
to zero.  This analysis clearly requires more complete data for $\meanB_\phi$,
however.
Given that a much longer data record is now available, it is surprising that
the analysis of Yoshimura (1976) has never been repeated;
better constraints on this quantity would be very useful.

In the following we emphasize the importance of determining the sign of
the magnetic helicity of the LS field and we suggest that, on theoretical
grounds, its sign should be opposite to that of the SS field.
We know that in active regions the magnetic helicity 
is negative (positive)
in the northern (southern) hemisphere (Seehafer 1990,
Pevtsov, Canfield \& Metcalf 1995, Bao et al.\ 1999, D\'emoulin et al.\ 2002).
We use this to support our expectation that what is observed in active
regions is actually the helicity of the SS field.

\section{RELATIONSHIP BETWEEN LARGE AND SMALL SCALE FIELDS}

Kinematic theory is able to explain the growth of helical fields
at large scales (LS) and the growth of helical and non-helical fields at
small scales (SS).
The growth rate of the SS field is usually much larger than that of the
LS field, and it was therefore thought that the rapid growth of this
SS fields tends to suppress the growth of the LS field
(Kulsrud \& Anderson 1992), and that the solar magnetic field may
therefore be of primordial origin (Vainshtein \& Cattaneo 1992).
This seemed however to be in conflict with several successful numerical
simulations of large scale dynamo action that showed the SS field to be merely
comparable in strength to the LS field (Glatzmaier \& Roberts 1995,
Brandenburg et al.\ 1995, Brandenburg 2001).

Meanwhile it has become clear that it is not the SS field as such that
causes problems, but only the helical part of the SS field that is generated
simultaneously with the LS field.
This phenomenon has been seen in simulations of MHD turbulence
with helical forcing (Brandenburg 2001) and has been
modeled successfully with semi-analytic, nonlinear two-scale  theories
(Field \& Blackman 2002; Blackman \& Brandenburg 2002; Blackman \& Field 2002).
These approaches assume (perhaps correctly)  
that the dynamo works primarily via helicity-dependent effects,
by which we mean both the $\alpha$-effect (Steenbeck, Krause, R\"adler 1966)
or the inverse cascade/transfer (Pouquet, Frisch \& L\'eorat 1976).
Some alternatives have been proposed: the Vishniac-Cho (2001) effect
(but see Arlt \& Brandenburg 2001) and negative turbulent magnetic diffusivity
(Zheligovsky, Podvigina \& Frisch 2001).

The essence of the nonlinearity of helical dynamos 
is that the helicity effects produce large scale helical magnetic
fields, but the total magnetic helicity is conserved, so there must
be a simultaneous generation of magnetic fields of opposite helicity.
Splitting the field into LS and SS components, $\BB=\meanBB+\bb$,
and likewise for the magnetic vector potential, $\AAA=\meanAA+\aaa$,
we have to satisfy
\EQ
\bra{\AAA\cdot\BB}=\bra{\meanAA\cdot\meanBB}+\bra{\aaa\cdot\bb}
\approx0.
\EN
If the LS field were fully helical, and of typical wavenumber $k_{\rm m}$,
we would have
$k_{\rm m}\bra{\meanAA\cdot\meanBB}=\pm\bra{\meanBB^2}$,
depending on the sign of the magnetic helicity of the LS field.
For fractional helicity of the forcing one has instead
\EQ
k_{\rm m}\bra{\meanAA\cdot\meanBB}=\epsilon_{\rm m}\bra{\meanBB^2},
\EN
where $|\epsilon_{\rm m}|\le1$, and $\epsilon_{\rm m}=0$ for
non-helical fields.
A similar relation applies to the fluctuating field (of wavenumber $k_{\rm f}$)
and its helicity,
i.e.\ $k_{\rm f}\bra{\aaa\cdot\bb}=\epsilon_{\rm f}\bra{\bb^2}$.
With these preliminaries we can say that in a magnetic helicity
conserving situation
\EQ
{\bra{\meanBB^2}\over\bra{\bb^2}}={k_{\rm m}\over k_{\rm f}}
\,{\epsilon_{\rm f}\over\epsilon_{\rm m}}.
\EN
If  $|\epsilon_{\rm f}|\approx|\epsilon_{\rm m}|$, and since
$k_{\rm m}<k_{\rm f}$ for finite scale separation, we have in
this phase $\bra{\meanBB^2}<\bra{\bb^2}$.
We note, however, that in $\alpha\Omega$ dynamos where a strong toroidal
LS field can be generated from a poloidal LS field, regardless of
helicity --- so that $k_{\rm m}\bra{\meanBB^2}\gg|\bra{\meanJJ\cdot\meanBB}|$ ---
one may have $|\epsilon_{\rm m}|\ll|\epsilon_{\rm f}|$.
It is then possible that $\bra{\meanBB^2}$ can be comparable to
or in excess of $\bra{\bb^2}$, or at least it helical component
(Blackman \& Brandenburg 2002).

To summarize, magnetic helicity conservation links the magnitude of
the helical components of the LS and SS fields in a well defined way,
depending mainly on the degrees to which LS and SS fields are helical,
i.e.\ on the values of $\epsilon_{\rm m}$ and $\epsilon_{\rm f}$.
For simple dynamos in slab geometry these values can be determined from
kinematic theory (Blackman \& Brandenburg 2002).

\section*{SIMULTANEOUS PRODUCTION OF LARGE AND SMALL SCALE TWIST}

\begin{figure}[t!]\begin{center}
\includegraphics[width=.4\textwidth]{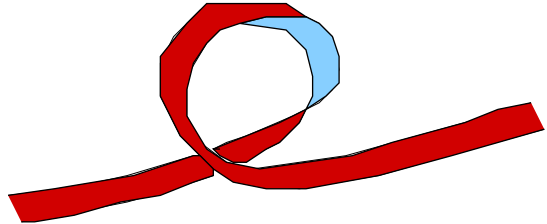}
\end{center}\caption[]{
Tilting of the rising tube due to the Coriolis force.
Note that the tilting of the rising loop causes also internal twist.
}\label{Fribbon2}\end{figure}

Given that magnetic helicity is conserved in the absence of boundary losses
and resistivity, any swirl-like motion must simultaneously introduce oppositely
helical magnetic fields when starting with an initially non-helical magnetic
field (Longcope \& Klapper 1997).
The prime example is of course the formation of an $\Omega$-shaped
flux loop due to magnetic or thermal buoyancy, and the simultaneous tilting
due to the Coriolis force. This is sketched in Fig.~\ref{Fribbon2}.

\begin{figure}[b!]\begin{center}
\includegraphics[width=.7\textwidth]{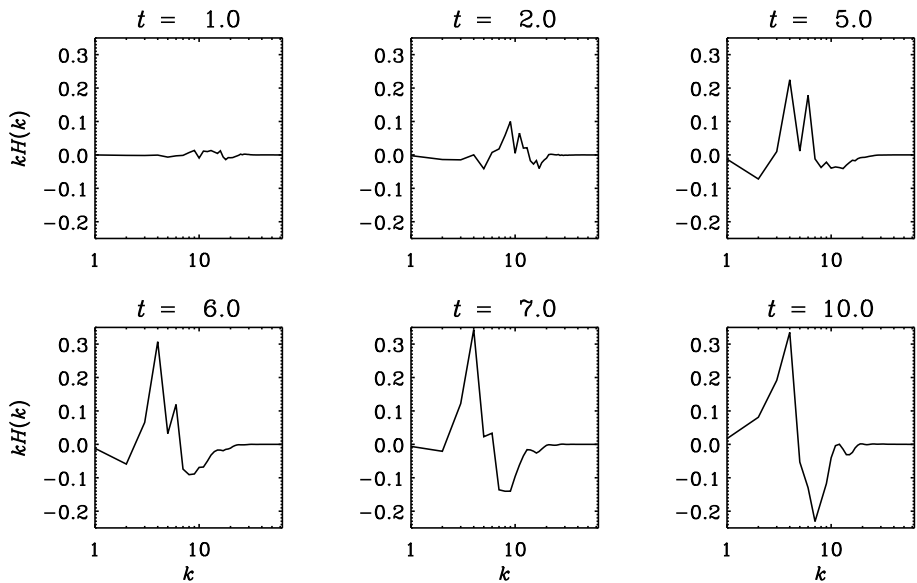}
\end{center}\caption[]{
Magnetic helicity spectra (scaled by wavenumber $k$ to give magnetic
helicity per logarithmic interval) taken over the entire computational
domain. The spectrum is dominated by a positive component at
large scales ($k=1-5$) and a negative component at small scales ($k>5$).
}\label{Fpspec}\end{figure}

We consider now the result of a simulation of a buoyant magnetic flux
tube. Similar calculations have been carried out many times in the past
(e.g.\ Abbett, Fisher \& Fan 2000),
but here we are interested in the magnetic helicity spectrum, which
seems to have attracted little attention so far. We start with a
horizontal flux tube in the azimuthal ($y$-) direction with vanishing
net flux (so there is a weak oppositely oriented field outside the tube)
and a $y$-dependent sinusoidal modulation of the entropy along the tube.
This destabilizes the tube such that it rises in one portion of the
box.
Although the box is not periodic in the vertical direction, the boundary
conditions are still sufficiently far away that we can obtain power spectra 
of the magnetic helicity via Fourier transforms; see Fig.~\ref{Fpspec}.
Note that after some time ($t=6$ free-fall times) the spectrum
begins to show mostly positive magnetic
helicity (as expected), together with a gradually increasing higher
wavenumber component of negative spectral helicity density.
The latter is the anticipated contribution from small scales
resulting from the twist of the tube.
%%ngrs: eyeballing e.g. the t=5 plot suggests that helicity hasn't been 
%%  conserved (?).  is the resistive time short cf. the freefall time (?!), 
%%  can these plots not be integrated (over log scales) for total helicity,
%%  or is my eyeball simply misleading?!
%AB: haven't investigated this yet.
%AB: Eric, any idea?

\begin{figure}[t!]\begin{center}
\includegraphics[width=.7\textwidth]{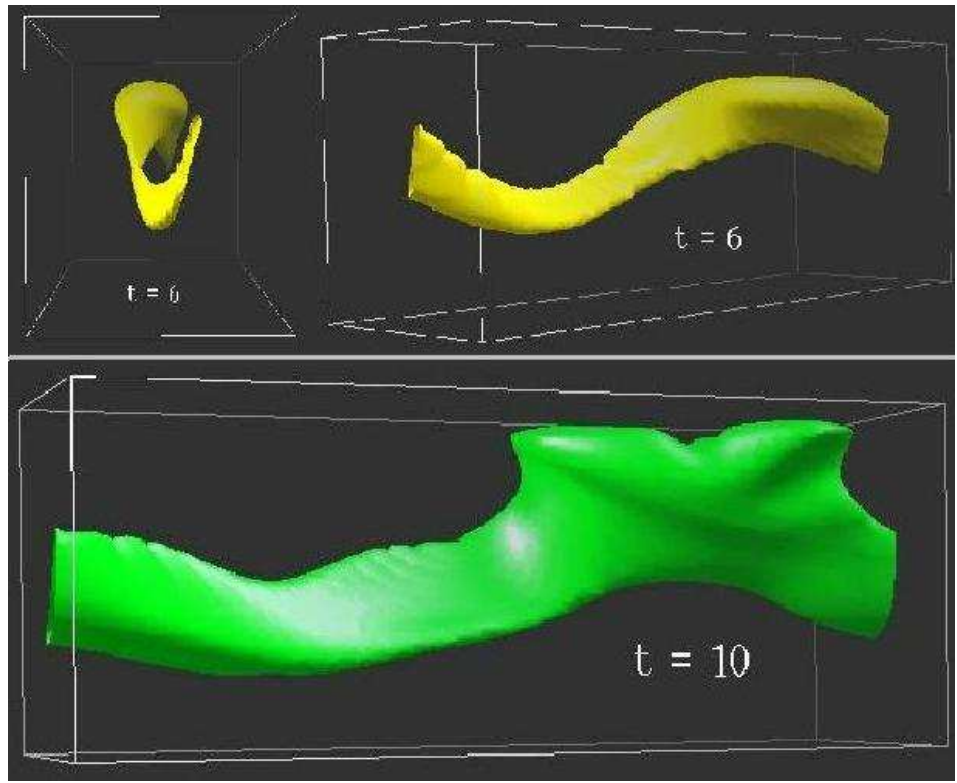}
\end{center}\caption[]{
Three-dimensional visualization of a rising flux tube in
the presence of rotation. The stratification is adiabatic such
that temperature, pressure, and density all vanish at a height
that is about 30\% above the vertical extent shown. (The actual
computational domain was actually larger in the $x$ and $z$ directions.)
}\label{Fall}\end{figure}

Instead of visualizing the magnetic field strength, which can be strongly
affected by local stretching, we visualize the rising flux tube using
a passive scalar field that was initially concentrated along the flux
tube. This is shown in Fig.~\ref{Fall}.

\section*{WHY SMALL SCALE LOSSES ARE GOOD}

A relatively useful concept is based on the evolution equations for
SS and LS fields under the assumption that the fields are
maximally helical (or have known helicity fractions $\epsilon_{\rm m}$
and $\epsilon_{\rm f}$) and have opposite signs of magnetic helicity at small
and large scales. The details can be found in Brandenburg, Dobler \&
Subramanian (2002, Sect.~4.2) and Blackman \& Brandenburg (2003).
The strength of this approach is that
it is quite independent of mean-field theory.

Losses of large-scale field have been modeled using diffusion terms.
The phenomenological evolution equation are written in terms of the
LS and SS magnetic energies, $M_{\rm m}$ and
$M_{\rm f}$ respectively, where we assume
$M_{\rm m}=\pm\mu_0 C_{\rm m}/k_{\rm m}$ and
$M_{\rm f}=\mp\mu_0 C_{\rm f}/k_{\rm f}$ for fully helical fields
(upper/lower signs apply to northern/southern hemispheres).
Here, $C_{\rm m}=\bra{\meanJJ\cdot\meanBB}$ and $C_{\rm f}=\bra{\jj\cdot\bb}$
are the LS and SS current helicities.
The phenomenological evolution equation for the LS energy then takes the form
\begin{equation}
k_{\rm m}^{-1}{{\rm d}M_{\rm m}\over{\rm d}t}=
-2\eta_{\rm m}k_{\rm m}M_{\rm m}
+2\eta_{\rm f}k_{\rm f}M_{\rm f},
\label{evolv_pheno}
\end{equation}
where $\eta_{\rm m}$ and $\eta_{\rm f}$ are effective magnetic
diffusivities that are expected to lie somewhere between between the molecular
magnetic diffusivity, $\eta$, and the turbulent magnetic diffusivity,
$\eta_{\rm t}$. 
%%ngrs:  the original of the following was rather tautological;  
%%  the following may be too trite, however.
%AB: will check tomorrow...
The positive sign for the term involving $M_{\rm f}$ reflects the generation 
of the LS field from the SS.
The case $\eta_{\rm m}=\eta_{\rm f}=\eta$ was already
discussed by Brandenburg (2001) who assumed that the small scale
magnetic field saturates at a certain time $t_{\rm sat}$, 
so that $M_{\rm f}\approx\mbox{const}$ for $t>t_{\rm sat}$. After that time,
Eq.~(\ref{evolv_pheno}) can be solved to give
\begin{equation}
M_{\rm m}=M_{\rm f}\,{\eta_{\rm f}k_{\rm f}\over\eta_{\rm m}k_{\rm m}}
\left[1-e^{-2\eta_{\rm m}k_{\rm m}^2(t-t_{\rm sat})}\right],
\quad\mbox{for $t>t_{\rm sat}$}.
\label{solution_pheno}
\end{equation}
This equation shows three things:
\begin{itemize}
\item
The time scale on which the large scale magnetic energy evolves
depends only on $\eta_{\rm m}$, not on $\eta_{\rm f}$.
\item
The saturation amplitude diminishes as $\eta_{\rm m}$ is
increased, which compensates the accelerated growth just past
$t_{\rm sat}$ (Brandenburg \& Dobler 2001).
\item
The reduction of the saturation amplitude due to $\eta_{\rm m}$
can be offset by having $\eta_{\rm m}\approx\eta_{\rm f}$, i.e.\
by having losses of small and large scale fields that are about
equally important.
\end{itemize}

\begin{figure}[t!]\centering\includegraphics[width=0.5\textwidth]{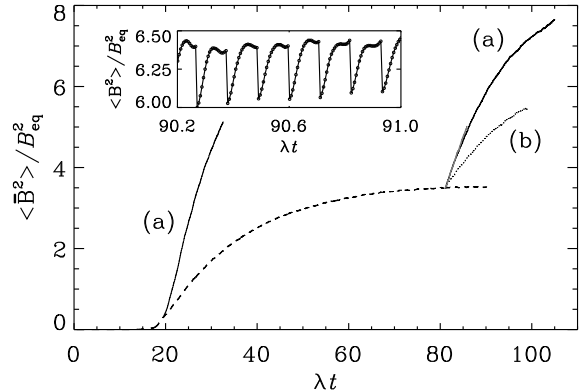}\caption{
The effect of removing small scale magnetic energy in regular time intervals
$\Delta t$ on the evolution of the large scale field (solid lines).
The dashed line gives the evolution of $\bra{\meanBB^2}$ for Run~3 of
Brandenburg (2001), where no such energy removal was included.
In all cases the field is shown in units of
$B_{\rm eq}^2=\mu_0\rho_0\bra{\uu^2}$. The two solid lines show the evolution of
$\bra{\meanBB^2}$ after restarting the simulation from Run~3 of B01
at $\lambda t=20$ and $\lambda t=80$. Time is scaled with the kinematic
growth rate $\lambda$. The curves labeled (a) give the result for
$\Delta t=0.12\lambda^{-1}$ and those labeled (b) for $\Delta t=0.4\lambda^{-1}$.
The inset shows, for a short time interval, the
sudden drop and subsequent recovery of the total (small and large scale)
magnetic energy in regular time intervals.
[Adapted from Brandenburg, Dobler \& Subramanian (2002).]
}\label{Fpbmean}\end{figure}

The overall conclusions that emerge are: (i) $\eta_{\rm m}>\eta$
is required if the large scale field is to evolve on a time scale other
than the resistive one; (ii) $\eta_{\rm m}\approx\eta_{\rm f}$
is required
if the saturation amplitude is not to be catastrophically diminished.
%[In fact, $\eta_{\rm m}\ga\eta_{\rm f}$ if we require the rates of
%LS and SS helicity loss to be the same; see Blackman \& Brandenburg (2003).]
These requirements are perfectly reasonable, but so far they have not been
borne out by simulations. Brandenburg \& Dobler (2001) found that most
of the losses of magnetic helicity occur on large scale. This is at first
glance very surprising, but on the other hand the magnetic helicity is a
quantity that is strongly dominated by the large scales. However,
certain phenomena such as CMEs and other perhaps less violent surface
events are not presently included in the simulations. As vindication of
the concept, however, it has been possible to show that the artificial
removal of small scale magnetic fields (via Fourier filtering after a
certain number of time steps) can indeed lead to significant increase
of the saturation amplitude (Brandenburg, Dobler \& Subramanian 2002).
This is shown in \Fig{Fpbmean}, where we show the evolution of the
LS magnetic energy in a run where, in certain time intervals, magnetic
energy is removed above the wavenumber $k=4$.
Comparison with the original curve of Run~3 of Brandenburg (2001)
shows that the removal of SS magnetic field allows the LS field to
grow beyond the original limit, which is in agreement with the
prediction from the phenomenology given by \Eq{solution_pheno}.
Furthermore, and again in agreement with \Eq{solution_pheno}, the
time scale for saturation is still the resistive time scale.

\section*{DISCUSSION}

After having discussed the relation between LS and SS helical fields, and
their significance to losses of helical fields through the
solar surface, we can now restate the reasons why we expect there to be
as yet undetected losses of large scale magnetic helicity of positive sign
in the northern hemisphere and negative sign in the southern hemisphere.
Because of magnetic helicity conservation, the observed magnetic helicity
of one sign must be accompanied by magnetic helicity of the opposite sign
(e.g. Blackman \& Field 2000). 
Simulations have shown that contributions of opposite sign occur at
different length scales, rather than at different positions in space.
This leaves two possibilities: there could be a change of sign of magnetic
helicity at a scale less than the resolution cutoff of about 80\km,
or there could be magnetic helicity at a scale comparable to the size
of the entire sun.
There are two argument in favor of the latter possibility.
(i) According to standard expectations, the $\alpha$-effect
should be positive (negative) in the northern (southern) hemisphere.
Since the $\alpha$-effect produces the field on the scale of the Sun,
the helicity of the latter should also be positive (negative) in the northern (southern)
hemisphere.
(ii) If $\alpha$ had the opposite sign (as found for example
in simulations of accretion disc dynamos; see Brandenburg et al.\ 1995),
the observed losses would be associated with large scale field.
Predominant large scale losses would however diminish the magnitude of
the observed large scale field, to a level probably far below
equipartition.

\section*{CONCLUDING REMARKS}

The magnetic helicity equation has proved to be a valuable tool
in understanding mean-field dynamos based on helicity effects.
This tool is most successful in connection with homogeneous dynamos,
where the kinetic helicity distribution is uniform.
Of great interest is of course the nonuniform case that has been
discussed in a number of recent papers by Kleeorin et al.\ (2000, 2002).
The problems that arise if one relaxes the restrictions to nonuniformity
can be traced back to the absence of gauge invariant formulations of
the helicity equation in that sense (Brandenburg 2003).
Because of these difficulties we have here adopted a more phenomenological
approach where the helicity gradient terms are simply modeled as diffusion
terms.

Simulations with open boundaries and stratification
are needed to check whether more realistic dynamos always exhibit bi-helical
behavior, as suggested in the present work.
This is also related to the question of the observational appearance of
bi-helical features.
In Blackman \& Brandenburg (2003) we have suggested that the {\sf N}
and {\sf S}-shaped sigmoidal structures could be a direct manifestation
of bi-helical structure.
Viewed from the top, the flux tube in Fig.~\ref{Fribbon2} would look
like a {\sf N}, in agreement with what is expected from Joy's law
for the norther hemisphere, and like a {\sf S} for the southern hemisphere.
It would be an important validation of a simulation of large scale dynamo
action if such sigmoidal structures could be obtained self-consistently.

\section*{ACKNOWLEDGEMENTS}

Use of the supercomputers in Odense (Horseshoe)
and Leicester (Ukaff) is acknowledged.

\end{document}